\documentclass{iopart}
\usepackage{graphicx}
\def\pt{$p_T$}
\def\rec{recombination}
\def\frag{fragmentation}

\begin{document}

\title{Fragmentation or Recombination at High $p_T$?}
\author{Rudolph C. Hwa$^1$ and  C.\ B.\ Yang$^{1,2}$}
\address{$^1$Institute of Theoretical Science and Department of
Physics\\ University of Oregon, Eugene, OR 97403-5203, USA\\
$^2$Institute of Particle Physics, Hua-Zhong Normal University,
Wuhan 430079, P.\ R.\ China}

\begin{abstract}
All hadronization processes, including fragmentation, are shown to proceed through
\rec. The shower partons in a jet turn out to play an important role in
describing the \pt\ spectra of hadrons produced in heavy-ion collisions.
Due to the \rec\ of the shower partons with the soft thermal partons,  the
structure of jets produced in $AA$ collisions is not the same as that of
jets produced in $pp$ collisions.
\end{abstract}

The question is: ``fragmentation or recombination?" The answer is not even
fragmentation {\it and} recombination, but \rec\   only.  What is usually
called \frag\ is not a description of the hadronization process; it is a
phenomenological parametrization of a black box represented by the function
$D(z)$ that gives the momentum fraction $z$ of a hadron in a parton jet. We
shall open up that black box and formulate the so-called \frag\ process in
terms of \rec. To do so, we need to determine the distributions of the shower
partons in a jet \cite{hy1}. Those shower partons can recombine among
themselves
to form hadrons that are usually regarded as the  products of the
\frag\ process,
but they can also recombine with the thermal partons not belonging to the jet.
The latter process has hitherto been overlooked, but turns out to be
very important
in the moderate \pt\ range, $2<p_T<9$ GeV/c \cite{hy2}.

The \frag\ function (FF) can be expressed in the \rec\ model as a
convolution \cite{hy1}
\begin{eqnarray}
zD_i^M(z)=\int{dx_1\over x_1}{dx_2\over x_2} F_{q\bar q'}^i(x_1,x_2)\
R^M(x_1,x_2,z) \ ,    \label{1}
\end{eqnarray}
where $M$ is the meson in a jet initiated by a hard parton $i$.
$F^i_{q\bar q'}(x_1,x_2)$
is the joint distribution of the shower partons $q$ and $\bar q'$ in
the jet, which are to
recombine to form $M$. The \rec\ function $R^M(x_1,x_2,z)$ is known
from previous studies
\cite{hw,hy3}. Since FFs are known from the parametrizations in
\cite{bkk}, we can
determine $F^i_{q\bar q'}$  from (\ref{1}).  Let $S_i^j$ denote the shower
parton distribution (SPD) of a parton $j$ in the shower initiated by a hard parton $i$.
On the assumption that
$F^i_{q\bar q'}$ are
factorizable into products of two SPDs, it
is possible to reduce
the 12 elements of the shower matrix $S_i^j$ from hard parton
$i=u,d,s,g$ to shower parton
$j=u,d,s$ to just 5 basic distributions, $K_{\rm NS}, L, G, L_s, {\rm
and}\  G_s$, as follows:
\begin{equation}
S_i^j=\pmatrix{K&L&L_s\cr
                   L&K&L_s\cr
                 L&L&K_s\cr
                 G&G&G_s\cr} \ ,
     \label{2}
\end{equation}
   where $K = K_{\rm NS} + L$ and $K_s = K_{\rm NS} + L_s$.
   An example of $K$ is $u\rightarrow u$ (valence + sea); $L$ can be
   $u\rightarrow d$ or $\bar d$, and $L_s$ can be $u\rightarrow s$.
   The five SPDs are shown in Fig.\ 1 as
   functions of their momentum fractions \cite{hy1}. The anti-quarks
$\bar{u}$, $\bar{d}$ and $\bar{s}$ have the same structure, and are
related to $u$, $d$, $s$ as sea, and vice-versa. The gluons are not
included among the shower partons $j$ because all gluons have been converted to quark pairs.
  Using those SPDs it is possible to calculate other FFs, both
measured (e.g., for proton)
  and unmeasured (e.g., for two pions).

\begin{figure}[tbph]
\includegraphics[width=0.65\textwidth]{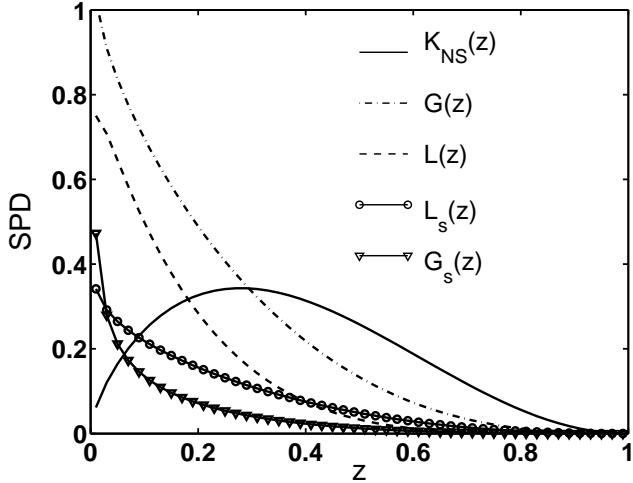}
\caption{Shower parton distributions plotted as functions of their
momentum fractions.}
\end{figure}

The above discussion on shower partons is independent of heavy-ion
collisions. However,
once the SPDs are known, their relevance to high-\pt\ hadrons
produced at RHIC is conceptually
unavoidable, since shower partons can recombine with thermal partons.
To make this point transparent,
consider the production of a $u$ quark at $p_T=8$ GeV/c by hard
scattering. Among the many shower
partons that it creates, we show in Fig.\ 2 by the solid line the distribution
of $u$ quarks in
the shower. That distribution corresponds to $K(z)$, which is the sum
of the valence
$K_{\rm NS}(z)$ and the sea $L(z)$ that are shown in Fig. 1.
The thermal partons are shown by the dashed line in Fig. 2. It should
be clear that the
recombination of the $u$  quark  in the shower with a $\bar d$
thermal quark to form a
$\pi^+$ cannot be ignored, when the parton momenta  are not too far
apart. The thermal-shower
\rec\ is important because the yield of that component is raised by
the high density of soft
partons, while the \pt\ of the produced hadrons is increased by
the semi-hardness of the
shower partons.

\begin{figure}[tbph]
\includegraphics[width=0.65\textwidth]{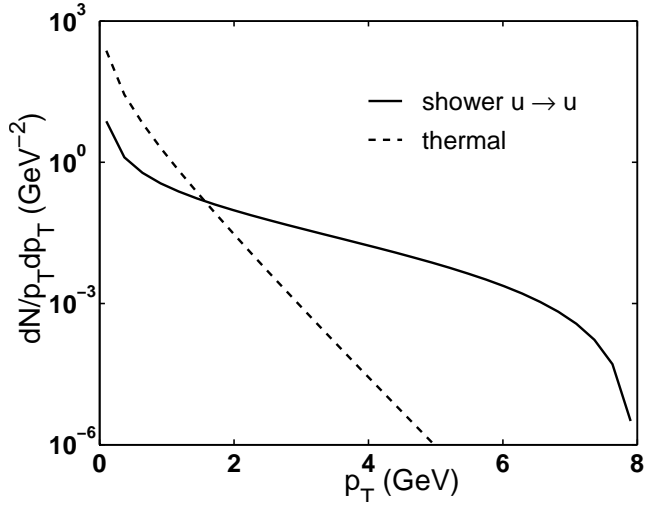}
\caption{
Shower parton distribution of $u$ quarks (in solid line) due to a
hard $u$ quark at 8 GeV/c.
The dashed line shows the thermal parton distribution.}
\end{figure}

The invariant distribution of pions in a 1D description is
\begin{eqnarray}
p{dN^{\pi}\over dp}=\int{dp_1\over p_1}{dp_2\over p_2} F_{q\bar
q'}^i(p_1,p_2)\ R^{\pi}(p_1,p_2,p) \ .
\label{2}
\end{eqnarray}
The $q\bar q'$ distribution has four components
\begin{eqnarray}
F_{q\bar q'}={\cal TT + TS + SS+(SS)}_2 \ ,   \label{3}
\end{eqnarray}
where $\cal T$ denotes thermal partons and $\cal S$ shower partons.
Thus ${\cal SS}$ represents two shower partons in one jet, and ${\cal
(SS)}_2$ signifies two shower
partons from two jets. A sum over all hard partons $i$ is implied.
Mathematical details of what each component
contains can be found in
Ref.\ \cite{hy2}. Letting $p$ in (3) represent $p_T$ and dividing that equation by
$p_T^2$ we can calculate $dN^{\pi}/p_Tdp_T$.  The result for the pion \pt\
spectrum in central Au+Au collisions at $\sqrt s=200$  GeV is shown in Fig.\ 3,
exhibiting good agreement with the
data \cite{ph}. Note that the thermal-shower component dominates over
other components in the
region $3<p_T<8$ GeV/c. The shower-shower (1 jet) component
(i.e., ${\cal SS}$) is the usual \frag\ and is not important until
$p_T>8$ GeV/c.

\begin{figure}[tbph]
\includegraphics[width=0.65\textwidth]{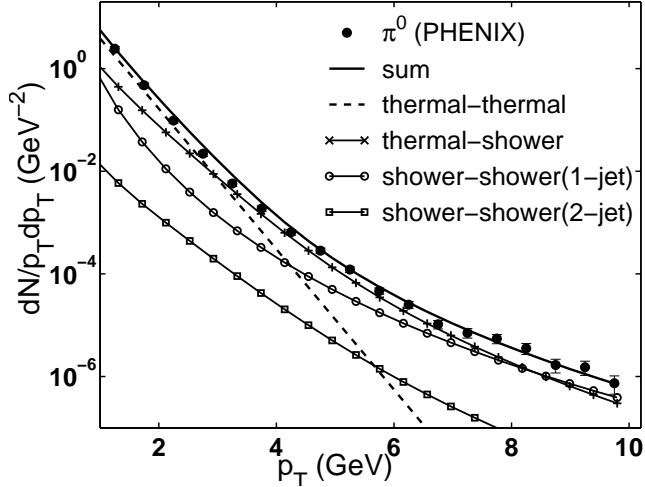}
\caption{Transverse momentum distribution of $\pi^0$ in Au-Au
collisions. Data are from \cite{ph}. The solid line is the sum of
four contributions to the recombination of partons: $\cal TT$ (dashed
line); $\cal TS$ (line with crosses); ${\cal SS}$, two shower
partons in one jet (line with open circles); ${\cal (SS)}_2$, two shower
partons from two overlapping jets (line with squares).}
\end{figure}

There is an essential parameter $\xi$ that has been adjusted to fit
the high-\pt\ data. It
represents the fraction of hard partons that can get out of the dense
medium created in central
collisions to produce the showers. It is an average quantity that
parametrizes the energy loss effect.
Thus $\cal S$ and $\cal SS$ are proportional to $\xi$, but not $\cal T$.
The value of $\xi=0.07$ is determined by fitting the
high-\pt\ spectrum. It is smaller
than the value of the nuclear modification factor,
$R_{AA}(p_T)\approx 0.2$, for $p_T>4$ GeV/c.
The reason for the difference is that the denominator of $R_{AA}$ is
the scaled $pp$ spectrum
(mainly \frag), which is smaller than our $\cal TS$ \rec, as is
evident in Fig.\ 3, the dominant
component that determines $\xi$.

It is important to emphasize that because of the dominance of
the $\cal TS$ \rec, the jet
structure in $AA$ collisions cannot be the same as that in $pp$
collisions, for which the
${\cal SS}$ \rec\ gives rise to the usual \frag. Thus there should be
more particles associated
with a trigger particle in $AA$ collisions than in $pp$ collisions
for the simple reason that the
abundance of soft thermal partons in the former case enhances the
hadrons formed at moderate \pt\
through \rec. There exists experimental evidence for the excess of
associated particles on the
same side, given in terms of the quantity $I_{AA}$ by STAR \cite{st}
for $3<p_T^{\rm trig}<4$
GeV/c. In that \pt\ range for the trigger particles it is found that
$I_{AA}\approx 1.5$,
whereas the value of 1 is expected if the jet structures in $AA$ and
$pp$ collisions are identical.

The formalism described above for Au+Au collisions has very recently
been extended to d+Au
collisions. It is found that the \rec\ of soft and shower partons can
account for the Cronin
effect \cite{cr} that is usually attributed to the broadening of the
parton $k_T$ width in the
initial state by multiple scattering \cite{aa}. In our study the
enhancement of the pion spectrum
at moderate \pt\ for more central d+Au collisions is due entirely to
the final-state interaction,
since there are more soft partons to recombine at higher centrality
\cite{hy4}. That is another
instance where the adherence to the usual \frag\ picture leads one to
a path that misses a crucial
element of the hadronization process.

The \rec\ model has by now been shown to explain many features of
hadron production in heavy-ion
collisions \cite{fr}. What we add here is a new piece of the \rec\
products, arising from the
interaction between the soft and shower partons. That new addition
has far-reaching consequences.
Theoretically, it means that the separation of partons into two
noninteracting components, soft
and hard, is an oversimplification. Experimentally, it means that
jets in $AA$ collisions are
different from those in $pp$ collisions. More precise quantification
of that difference can
provide further elucidation of the underlying hadronization process.

        This work was
supported, in part,  by the U.\ S.\ Department of Energy under
Grant No. DE-FG03-96ER40972 and by the Ministry of Education of
China under Grant No. 03113.

\end{document}